\documentclass[10pt,conference]{IEEEtran}
\usepackage{afterpage}
\usepackage{cite}
\usepackage{amsmath,amssymb,amsfonts}
\usepackage{algorithmic}
\usepackage{graphicx}
\usepackage[utf8]{inputenc}
\usepackage{listings}
\usepackage{multirow}
\usepackage{textcomp}
\usepackage{url}
\usepackage{xcolor}

\makeatletter

\def\RQOne{How do ML developers perceive and tackle the design of ML application systems and software?}
\def\RQTwo{How do academic and gray literatures address the design of ML application systems and software?}
\def\RQThree{How can ML architecture and design patterns be classified?}
\def\RQFour{What software-engineering architecture and design patterns for ML application systems and software exist?}
\newcommand{\answer}[2]{~\\[-0.4\baselineskip]\noindent\fbox{\parbox{\columnwidth}{#1 \textbf{#2}}}}

\begin{document}

\title{Studying Software Engineering Patterns for Designing Machine Learning Systems}

\author{\IEEEauthorblockN{Hironori Washizaki}
\IEEEauthorblockA{Waseda University \\
Tokyo, Japan \\
washizaki@waseda.jp}
\and
\IEEEauthorblockN{Hiromu Uchida}
\IEEEauthorblockA{Waseda University \\
Tokyo, Japan \\
eagle\_h.21@toki.waseda.jp}
\and
\IEEEauthorblockN{Foutse Khomh}
\IEEEauthorblockA{Polytechnique Montréal \\
Montréal, QC, Canada \\
foutse.khomh@polymtl.ca}
\and
\IEEEauthorblockN{Yann-Gaël Guéhéneuc}
\IEEEauthorblockA{Concordia University \\
Montréal, QC, Canada \\
yann-gael.gueheneuc@concordia.ca}
}

\maketitle

\begin{abstract}
Machine-learning (ML) techniques have become popular in the recent years. ML techniques rely on mathematics and on software engineering. 
Researchers and practitioners studying best practices for designing ML application systems and software to address the software complexity and quality of ML techniques. Such design practices are often formalized as architecture patterns and design patterns by encapsulating reusable solutions to commonly occurring problems within given contexts.
However, to the best of our knowledge, there has been no work collecting, classifying, and discussing these software-engineering (SE) design patterns for ML techniques systematically.
Thus, we set out to collect good/bad SE design patterns for ML techniques to provide developers with a comprehensive and ordered classification of such patterns. We report here preliminary results of a systematic-literature review (SLR) of good/bad design patterns for ML. 
\end{abstract}


\section{Introduction}

Machine-learning (ML) techniques have become popular in the recent years. They are being used in many domains, 
including cyber security, bioinformatics, IoT and autonomous cars. 
%\YANN{Foutse, could you add some well-known domains with some references?}
ML techniques rely on mathematics and on software engineering. They rely on mathematics for their algorithms and their capabilities to learn from input data and produce representative models. They also rely on software engineering for their implementations and their performances.

While there have been many works on the mathematics and computer science on which ML techniques are built, there have been fewer works on their implementations, which raises many concerns. First, users are concerned by the software complexity of using the techniques. Second, they are concerned about the quality of the available implementations, including performance and reliability. Finally, users are concerned about the quality of the models that could be negatively impacted by a software bug.

These concerns could be alleviated if developers could convince their users of the software quality of their implementations of the ML techniques. Consequently, researchers and practitioners studying best practices for designing ML application systems and software to address the software complexity and quality of ML techniques. Such design practices are often formalized as architecture patterns and design patterns by encapsulating reusable solutions to commonly occurring problems within given contexts in ML application systems and software design. 
%These works pertain to \YANN{Foutse, Hiro, could you add some academic references here?}. 

There are also many resources available on the Internet discussing various ML techniques and their concrete uses, from putting together a pipeline to implementing a Markov Decision Process. These pieces of gray literature are useful to developers putting together a ML systems by providing many examples and discussing many good/bad design patterns.

However, to the best of our knowledge, there has been no work collecting, classifying, and discussing these software-engineering (SE) design patterns for ML techniques systematically although such patterns could greatly help software developers in putting in place ML techniques for their users.

Thus, we set out to collect good/bad SE design patterns for ML techniques to provide developers with a comprehensive and ordered classification of such patterns. We report here preliminary results of a systematic-literature review (SLR) of good/bad design patterns for ML. We focus in this paper on (1) our method, (2) architecture and design (anti-)patterns, and (3) preliminary, quantitative results. We also report on ML developers' understanding of the techniques implementations.

We thus answer the following research questions:

\begin{itemize}[
    \setlength{\IEEElabelindent}{0pt}
    \setlength{\itemindent}{2\parindent}
    \setlength{\listparindent}{\parindent}]
\item[RQ1.] \RQOne{} We conducted a questionnaire-based survey to understand. This survey shows that ML engineers have little knowledge of the architecture and design patterns that could help them developing their ML application systems and software. 

\item[RQ2.] \RQTwo{} We conducted a SLR of both academic and gray literature and identified 10 academic papers and 28 gray documents, which we analysed to extract patterns.

\item[RQ3.] \RQThree{} We distinguish SE patterns for ML (such as patterns for designing ML application software) and non-SE patterns for ML (such as patterns for designing ML models) by analysing the content of the found documents. We classify these SE patterns with respect to two processes: the typical ML pipeline process and the typical SE process from ISO/IEC/IEEE 12207.

\item[RQ4.] \RQFour{} From our collection of patterns, we show that SE patterns for ML do exist and related to different phases of the software-development process and ML pipeline. We also give some examples of such patterns.
\end{itemize}

The rest of this paper is as follows: Section \ref{Section: Related Work} summarises the related work. Section \ref{Section: Survey} describes our survey and answers RQ1. Section \ref{Section: SLR} presents our SLR and a subset of its results. Section \ref{Section: Discussions} discusses our results. Section \ref{Section: Conclusion} concludes.

\section{Related Work}
\label{Section: Related Work}

There are surveys on general architecture and design patterns,  e.g., \cite{DBLP:conf/europlop/AvgeriouZ05,DBLP:journals/jss/AmpatzoglouCS13,DBLP:journals/jss/MayvanRY17}, mostly for object-oriented design. Surveys on architecture and design patterns exist for specific domains, e.g., multi-agent systems \cite{DBLP:books/sp/14/JuziukWH14}, IoT \cite{DBLP:conf/icse/WashizakiYHKKOO19}, or security \cite{DBLP:journals/ijsse/PondeS16}.

For the domain of software engineering for ML applications, case studies, practices, and patterns are available as independent documents. To the best of our knowledge, this is the first comprehensive survey on ML architecture and design patterns.

\section{Survey}
\label{Section: Survey}

Machine-learning techniques are concrete solutions to practical problems. Hence, ML developers may already have built a body of knowledge on the good/bad design practices of ML development. We ask RQ1.\ \RQOne{}

To answer RQ1, we asked 760+ developers at Japanese companies through Japan-wide mailing lists to developers and direct contact with developers during a workshop on software engineering. Developers answered anonymously the following three question regarding SE patterns for ML systems. In these questions, we distinguish between ``patterns'', which we define as some ``formal practices'' in any structured form, and ah-hoc practice that are suggestions or lore suggested by some developers but without any kind of formal form, e.g., suggestions extracted from a blog text.

\begin{itemize}[
    \setlength{\IEEElabelindent}{0pt}
    \setlength{\itemindent}{2\parindent}
    \setlength{\listparindent}{\parindent}]
\item[SQ1.] Do you refer to existing reference architectures or architectural patterns to design your own ML systems?

\item[SQ2.] How did you acquire and elicit requirements on ML systems? Patterns or practices? Any template or process? Or ad-hoc?

\item[SQ3.] How did you ensure non-functional features of ML systems? Patterns or practices? Any process? Or ad-hoc?
\end{itemize}

Out of the 760+ contacted participants, nine answered at least the first question for a response rate of 1\%. We expected such a low rate of answers because we contacted mostly developers who do not work with ML systems.

\begin{table}[ht]
\centering
\caption{Summary of the Survey Results}
\label{Table: Survey Results}
\begin{tabular}{|c|c|c|c|}
\hline
\multirow{2}{*}{SQ1.} &  \multicolumn{2}{|c|}{Yes} & No \\
\cline{2-4}
& \multicolumn{2}{|c|}{3 (General architecture, design and cloud patterns)} & 5 \\
\hline
\multirow{2}{*}{SQ2.} & Patterns & Template or Process & Ad-hoc \\
\cline{2-4}
& 0 & 2 & 7 \\
\hline
\multirow{2}{*}{SQ3.} & Patterns & Process & Ad-hoc \\
\cline{2-4}
& 1 & 1 & 6 \\
\hline
\end{tabular}
\end{table}

Table \ref{Table: Survey Results} summarises the results of the surveys. The number of answer per columns differ because some participants only answered one or two of the three questions. It shows that most of the developers use ad-hoc practices. When they use patterns, developers mention that they use general patterns, like the design patterns by Gamma et al.\ \cite{GoF94-Patterns} rather than patterns dedicated to ML development.

Thus, we conclude the survey as follows:

\answer{RQ1.\ \RQOne{}}{Developers use either general patterns or ad-hoc patterns and need patterns dedicated to SE design of ML application systems and software.}

\section{Systematic Literature Review}
\label{Section: SLR}

\subsection{Queries}

We perform a systematic-literature review of both academic and gray literatures to collect software-engineering good/bad design patterns for ML application systems and software. For the academic literature, we choose Engineering Village as meta-search engine and, on August $14^{th}$, 2019, with the query:

\begin{lstlisting}
((((system) OR (software)) AND (machine learning) AND ((implementation pattern) OR (pattern) OR (architecture pattern) OR (design pattern) OR (anti-pattern) OR (recipe) OR (workflow) OR (practice) OR (issue) OR (template))) WN ALL) + ((cpx OR ins OR kna) WN DB) AND (({ca} OR {ja} OR {ip} OR {ch}) WN DT)    
\end{lstlisting}

For the gray literature, we use the Google search engine, on August $16^{th}$, 2019, with the queries: 

\begin{lstlisting}
(system OR software) "Machine learning" (pattern OR "implementation pattern" OR "architecture pattern" OR "design pattern" OR anti-pattern OR recipe OR workflow OR practice OR issue OR template)
\end{lstlisting}

\noindent and:

\begin{lstlisting}
"machine implementation pattern" OR "architecture pattern" OR "design pattern" OR anti-pattern OR recipe OR workflow OR practice OR issue OR template
\end{lstlisting}

We thus retrieved 23 academic scholarly papers (S). 
Then, we performed a snowballing and obtained 9 additional scholarly documents (A). We searched the Internet using Google Search Engine and obtained 48 gray literature documents. 

For each of the documents, two of the authors vetted whether they should be included in our SLR or not. We started from the titles and abstracts and then read the entire documents to decide whether or not a document pertained to software-engineering practices for ML systems. By the very definitions of our queries, we kept 10 scholarly papers among 23. And we kept 25 gray documents, among which 6 were actual papers or books not (yet) referenced in Engineering Village and which we added to the set A. We call these papers, books, and documents ``documents'' in the following. We kept 3 additional scholarly documents among 9. Finally, we have 10 papers s1--s10 \cite{Shams18,DBLP:conf/nips/SculleyHGDPECYC15,DBLP:conf/hpca/WuBCCCDHIJJLLLQ19,DBLP:journals/corr/abs-1903-00278,DBLP:conf/safecomp/KlasV18,Ahmed2010SoftwareAO,DBLP:conf/lrec/BethardOB14,DBLP:conf/caise/NalchigarYOCGC19,Liu18,DBLP:conf/icsa/Yokoyama19} in the set S, 19 (=25-6) documents g01--g19 \cite{g01,g02,g03,g04,g05,g06,g07,g08,g09,g10,g11,g12,g13,g14,g15,g16,g17,g18,g19} in the set G, and 9 (=3+6) documents a01--a09 \cite{DBLP:conf/vl/HilllazBEB16,DBLP:conf/ifip13/SeymoensOJVA18,DBLP:conf/osdi/LiAPSAJLSS14,DBLP:journals/corr/abs-1906-07154,DBLP:conf/icse/AmershiBBDGKNN019,DBLP:conf/iccS/NguyenCMA16,DBLP:journals/corr/SmithT16,Gollapudi16,Basak17} in the set A. All the data is available on-line\footnote{http://www.washi.cs.waseda.ac.jp/iwesep19/}.

\subsection{RQ2.\ \RQTwo{}}

\begin{figure}[ht]
\centering
\includegraphics[width=\columnwidth]{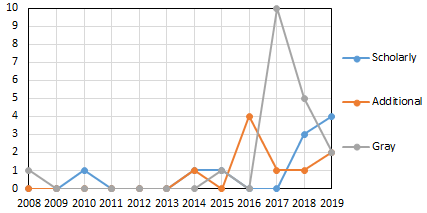}
\caption{Numbers of Documents per Year with S: Scholarly Papers, A: Additional Scholarly Documents (Snowballing), and G: On-line, Gray Documents}
\label{Figure: Trend Numbers per Year}
\end{figure}

Through our SLR, its process and results, we observed that ML systems are very popular thanks to the promotion of artificial intelligence in very recent years. Figure \ref{Figure: Trend Numbers per Year} shows the trend in the number of documents related to SE design for ML systems in the past 10 years.

ML systems are discussed in various on-line communities, from mathematics to library builders to the Maker Culture. The SE development of ML systems is more rarely the subject of academic research. In the gray literature, we found many documents discussing ML systems, often from the data scientists' points of view.

\answer{RQ2. \RQTwo{}}{There a few academic documents related to SE patterns for ML systems design. There are many gray documents discussing good/bad practices of ML systems design, at different levels of abstraction, focusing mostly on data management.}

\subsection{RQ3.\ \RQThree{}}

Through our SLR and while reading the documents, we took note of various characteristics that could help classify the patterns. We observed that in the academic and gray literatures, SE patterns for ML systems are often presented in the context of the ML pipeline or the SE development process. 

Consequently, we identified a general pipeline and development process in which each identified architectural and design pattern could fit. As pipeline, we chose the one presented by Microsoft \cite{DBLP:conf/icse/AmershiBBDGKNN019} composed by the nine stages: Model Requirements, Data Collection, Data Cleaning, Data Labelling, Feature Engineering, Model Training, Model Evaluation, Model Deployment, and Model Monitoring. 
For software development process, we chose the software implementation processes specified in ISO/IEC/IEEE 12207:2008 standard \cite{ISO12207} as follows: Requirements Analysis, Architectural Design, Detailed Design, Construction, Integration, and Qualification Testing. These pipelines and processes describe the stages and phases expected for developing and running any ML application systems, independent of its purpose and domain.

%\begin{figure*}[ht]
%\centering
%\includegraphics[width=\textwidth]{"ML Pipeline (Microsoft)".png}
%\caption{ML Pipeline from Microsoft \cite{DBLP:conf/icse/AmershiBBDGKNN019}}
%\label{Figure: ML Pipeline}
%\end{figure*}

%\afterpage{
%\begin{figure}
%\centering
%\includegraphics[width=.7\columnwidth]{"SE Process (ISO12207)".png}
%\caption{SE Process from ISO 12207 (excerpt\protect\footnotemark)}
%\label{Figure: SE Process}
%\end{figure}
%}
%\footnotetext{\url{https://msritse2012.wordpress.com/2013/01/30/isoiec-12207software-life-cycle-process/}}

Thus, we answer the question as follows:

\answer{RQ3. \RQThree{}}{SE patterns for ML systems divide along two main dimensions: ML pipeline and SE development process.}

\subsection{RQ4.\ \RQFour{}}

Two of the authors then read half of the documents each and extracted patterns while the other author vetted each pattern independently. We thus identified 69 patterns. Then, each other studied the others' patterns to retain only the 33 patterns related to the architecture and design of ML systems. 

Table \ref{Table: List of ML patterns} shows the list of extracted ML (anti-)architecture and design patterns: 18 (55\%) patterns have been extracted from the scholarly papers and additional scholarly documents, while 15 (45\%) from the gray documents. 

Then, we classified each pattern according to the stages of the ML pipeline and phases of the SE development processes.
%in Figures \ref{Figure: ML Pipeline} and \ref{Figure: SE Process}. 
Table \ref{Table: RQ4 Classification} shows our classification of the identified patterns. It shows that architecture patterns and design patterns pertain, by their very definition, mostly to the phase of Architectural Design, Detailed Design, and Construction but also to later phases. It shows also that they are mostly concerned with the later stages of the ML pipeline. We explain this observation by our choice of focusing on architecture patterns and design patterns that have mostly an impact on later stages.

\begin{table}[ht]
\centering
\caption{Extracted Patterns (``Anti?" denotes anti-patterns)}
\label{Table: List of ML patterns}
\begin{tabular}{|c|c|p{4.2cm}|c|}
\hline
Source & ID & Pattern Name & Anti? \\
\hline
\cite{DBLP:conf/nips/SculleyHGDPECYC15} & s02a & Glue Code & Y \\
\cite{DBLP:conf/nips/SculleyHGDPECYC15} & s02b & Wrap Black-Box Packages into Common APIs & \\
\cite{DBLP:conf/nips/SculleyHGDPECYC15} & s02c & Pipeline Jungles & Y \\
\cite{DBLP:conf/nips/SculleyHGDPECYC15} & s02d & Design Holistically about Data Collection and Feature Extraction & \\
\cite{DBLP:conf/nips/SculleyHGDPECYC15} & s02e & Dead Experimental Codepaths & Y \\
\cite{DBLP:conf/nips/SculleyHGDPECYC15} & s02f & Reexamine Experimental Branches Periodically & \\
\cite{DBLP:conf/nips/SculleyHGDPECYC15} & s02g & Abstraction Debt & Y \\
\cite{DBLP:conf/nips/SculleyHGDPECYC15} & s02h & Parameter-Server Architecture & \\
\cite{DBLP:conf/nips/SculleyHGDPECYC15} & s02i & Plain-Old-Data Type Smell & Y \\
\cite{DBLP:conf/nips/SculleyHGDPECYC15} & s02j & Descriptive Data Type for Rich Information & \\
\cite{DBLP:conf/nips/SculleyHGDPECYC15} & s02k & Multiple-Language Smell & Y \\
\cite{DBLP:conf/nips/SculleyHGDPECYC15} & s02l & Undeclared Consumers & Y \\
\cite{DBLP:conf/hpca/WuBCCCDHIJJLLLQ19} & s03a & Decouple Training Pipeline from Production Pipeline & \\
\cite{DBLP:conf/hpca/WuBCCCDHIJJLLLQ19} & s03b & ML Versioning & \\
\cite{DBLP:conf/safecomp/KlasV18} & s05 & Isolate and Validate Output of Model & \\
\cite{DBLP:conf/icsa/Yokoyama19} & s10a & Distinguish Business Logic from ML Models & \\
\cite{DBLP:conf/icsa/Yokoyama19} & s10b & Gateway Routing Architecture & \\
\hline
\cite{DBLP:journals/corr/abs-1906-07154} & a04 & Separation of Concerns and Modularization of ML Components & \\
\hline
\cite{g02} & g02a & Federated Learning & \\
\cite{g02} & g02b & Secure Aggregation & \\
\cite{g05} & g05 & Handshake or Hand Buzzer & \\
\cite{g07} & g07a & Test Infrastructure Independently from ML & \\
\cite{g07} & g07b & Reuse Code between Training Pipeline and Serving Pipeline & \\
\cite{g08} & g08 & Data-Algorithm-Serving-Evaluator & \\
\cite{g09} & g09 & Closed-Loop Intelligence & \\
\cite{g10} & g10 & Canary Model & \\
\cite{g12} & g12 & Daisy Architecture & \\
\cite{g13} & g13 & Event-driven ML Microservices & \\
\cite{g15} & g15a & Big Ass Script Architecture & Y\\
\cite{g12,g15} & g15b & Microservice Architecture & \\
\cite{g14,g16,Gollapudi16} & g16 & Data Lake & \\
\cite{g17} & g17 & Kappa Architecture & \\
\cite{g14,g18,Basak17} & g18 & Lambda Architecture & \\
\hline
\end{tabular}
\end{table}

\begin{table*}[ht]
\centering
\caption{Classification of the Identified Patterns}
\label{Table: RQ4 Classification}
\begin{tabular}{|p{1cm}|p{1cm}|c|c|c|c|p{1cm}|p{2cm}|p{4cm}|p{2cm}|}
\hline
     & \rotatebox{90}{Model Requirements~} & \rotatebox{90}{Data Collection~} & \rotatebox{90}{Data Cleaning~} & \rotatebox{90}{Data Labelling~} & \rotatebox{90}{Feature Engineering~} & \rotatebox{90}{Model Training~} & \rotatebox{90}{Model Evaluation~} & \rotatebox{90}{Model Deployment~} & \rotatebox{90}{Model Monitoring~} \\
\hline
\multicolumn{1}{|l|}{Requirements Analysis} & & & & & & a04 & a04 & a04, g13 & a04, g13 \\
\hline
\multicolumn{1}{|l|}{Architectural Design} & g02a, g02b, g05, g09 & & & & & s03a, a04, g02a, g02b & g07a, a04 & g07a, g07b, s03a, s05, s10b, a04, s02a, s02b, s02c, s02d, s02e, s02f, g10, g13, g02a, g02b, g05, g08, g09, g15a, g15b, g16, g17, g18, s10a & s05, s10b, a04, g10, g09, g13, g16, g17, g18 \\
\hline
\multicolumn{1}{|l|}{Detailed Design} & & & & & & & s02e, s02f, g10 & g07b, s02a, s02b, s02i, s02j, g10, g13 & s02i, s02j, g13 \\
\hline
\multicolumn{1}{|l|}{Construction} & & & & & & & s02e, s02f & g07b, s02a, s02b, s02c, s02d, s02e, s02f, s02g, s02h, s02i, s02j, s02k, g13 & s02i, s02j, g13 \\
\hline
\multicolumn{1}{|l|}{Integration} & & & & & & s03a & s02e, s02f & g07b, s03a, s05, s10b, s02e, s02f, s02l, g10 & s05, s10b, s02l, g10 \\
\hline
\multicolumn{1}{|l|}{Qualification Testing} & & & & & & & g07a, s02e, s02f & g07a, s03b, s02e, s02f & s03b \\
\hline
\end{tabular}
\vspace*{-0.5cm}
\end{table*}

Thus, we answer the question as follows:

\answer{RQ4. \RQFour{}}{There exist architecture patterns and design patterns. Some patterns apply to many stages of the pipeline, some to many phase of the development process, while others only apply to one stage and one phase only.}

\section{Discussions}
\label{Section: Discussions}

We describe succinctly two extracted patterns. We omit for the sake of brevity: participants, collaborations, implementation, sample code, and known uses. We then discuss threats to the validity of our results.

\subsection{Example of Architectural Pattern}

\paragraph{Pattern Name} s10a: Distinguish Business Logic from ML Model (originally named as ``Multi-Layer Architectural Pattern" \cite{DBLP:conf/icsa/Yokoyama19})

\paragraph{Intent} Isolate failures between business logic and ML learning to help developers debug ML application systems.

\paragraph{Also Known As} Machine Learning System Architectural Pattern for Improving Operational Stability.

\paragraph{Motivation} ML application systems are complex systems because their ML components must be (re)trained regularly and have a non-deterministic behaviour by nature. Business requirements for these systems, as any other systems, will also change as well as ML algorithms.

\begin{figure}[ht]
\centering
\includegraphics[width=9cm]{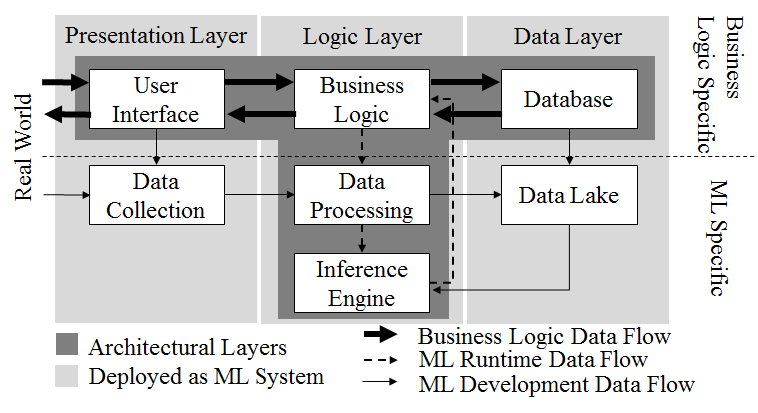}
\vspace*{-0.5cm}
\caption{Structure of Distinguish Business Logic from ML Model pattern \cite{DBLP:conf/icsa/Yokoyama19}}
\label{Figure: Architecture Pattern Example}
\vspace*{-0.5cm}
\end{figure}

\paragraph{Structure} Define clear APIs between traditional and ML components. Put business and ML components having different responsibilities into three layers as shown in Figure \ref{Figure: Architecture Pattern Example}. Divide data flows into three data flows.

% Deploy these components on separate computers (or virtual machines). Implement loggers and monitor the resulting logs to identify anomalies in the ML components (or their consequences in the business components).

\paragraph{Applicability} Any ML application systems whose output depends on ML techniques.

\paragraph{Consequences} Decoupling ``traditional'' business and ML components allows monitoring and adjusting the ML components to users' requirements and to changing inputs.

\subsection{Example of Design Anti Pattern}

\paragraph{Pattern Name} s02c: Pipeline Jungle \cite{DBLP:conf/nips/SculleyHGDPECYC15}

\paragraph{Intent} Maintain one controllable, straightforward pipeline of ML components.

\paragraph{Motivation} ML application systems combine several ML components with different input and output formats. These components in turn interact with business components. 

\paragraph{Problem} ML application systems may include ``glue code'' to scrape, join, and sample input/output data into one pipeline. This pipeline is fragile and must be maintained and tested carefully. Testing requires expensive end-to-end integration tests. Glue code is a technical debt that can prevent further innovations.

\paragraph{Refactored Solution} Define unit and component tests. If possible, convert input/output files into first-class objects and glue code into clear APIs.

\paragraph{Applicability} Any ML application systems using different techniques.

\paragraph{Related Patterns} s02a: Avoid Glue Code, s02b: Wrap Black-Box Packages into Common APIs, s02d: Design Holistically about Data Collection and Feature Extraction

\subsection{Threats to Validity}

As any empirical study, ours is subject to threats to its validity, which we summarise in the following.

\paragraph{Survey} Survey have well-known threats to their construct, internal, external, and conclusion validity. It is possible that by construct our survey does not ask relevant or answerable questions. We showed the questions in Section \ref{Section: Survey} and believe them both relevant and answerable. 

Internally, the questions of our survey could be contradictory or misleading. Again, by showing our questions in Section \ref{Section: Survey}, we try to alleviate this problem. Externally, our questions and their answers could be not generalizable to other participants or other domains. Our questions were general and did not assume any particular domain.

Finally, we could draw the wrong conclusion from the survey answers but backed all our answers by the data.

\paragraph{SLR} SLRs can also \cite{7890583} have threats to the validity of their results: mostly internal validity and reliability. Internal validity concerns the cause--effect conclusion that we drew from the SLR process and its results. We provided evidence from the data for each of the research question. 

Reliability concerns the quality and rigour with which we carried the SLR. We explained the different steps in Section \ref{Section: SLR} and reported the different number of documents at each step. Besides, we put all the data available on-line.

A threat to reliability is that an independent third-party has not vetted all the patterns that we identified. We intend to participate to Writers' Workshops at the Pattern Languages of Programs (PLoP) conference series\footnote{https://hillside.net/conferences/} to receive the community's feedback on each pattern before publishing them.

\section{Conclusion}
\label{Section: Conclusion}

In this paper we collected, classified, and analysed software-engineering architectural and design (anti-)patterns for machine-learning systems. We thus want to bridge the gap between traditional software systems and ML systems when it comes to their software architecture and design. We answered four research questions through a survey of software developers of ML systems and a systematic-literature review of both academic and gray literatures pertaining to ML systems and their software development.

RQ1 showed that SE developers are concerned by the complexity of ML systems and their lack of knowledge of the architecture and design (anti-)patterns that could help them. RQ2 returned more documents in the gray literature than in the academic literature. RQ3 showed that SE patterns for ML systems divide along two processes: ML pipeline and SE development process. RQ4 showed that SE patterns for designing ML systems exist and reported some examples.

As future work, we will complete our classification of the SE patterns for ML systems and produce a map of these patterns. We include non-architecture/non-design (anti-)pattern, such as SE patterns for requirements engineering of ML systems. We will study the impact of SE patterns on the quality attributes of ML systems. We will submit the patterns to the Writers' Workshop at the PLoP conference series.

\section*{Acknowledgement}

The authors would like to thank Prof. Naoshi Uchihira, Mr. Norihiko Ishitani, Dr. Takuo Doi, Dr. Shunichiro Suenaga and Mr. Yasuhiro Watanabe for their helps.
This work was supported by JST-Mirai Program Grant Number JP18077318, Japan.

\bibliographystyle{IEEEtran}
\bibliography{Main}

\end{document}